\documentclass[12pt]{article}
\usepackage{amsmath}
\usepackage{graphicx}
\date{}
\begin{document}
\begin{titlepage}
\Large
\title{Classical representation for quantum states of a particle  in $\lambda z^{2m}$ potential}
\author{\large
T.P. Grozdanov$^1$ and E.A. Solov'ev$^2$  \\
\normalsize $^1$Institute of Physics, University of Belgrade, Pregrevica 118, 11080 Belgrade, Serbia\\
\normalsize  Email: tasko@ipb.ac.rs\\
\normalsize $^2$Bogoliubov Laboratory for Theoretical Physics, Joint Institute for Nuclear Research,\\
\normalsize Dubna, 141980 Moscow Region, Russia\\
\normalsize  Email: esolovev@theor.jnr.ru\\
}
\end{titlepage} \maketitle
\begin{abstract}
A classical representation for quantum eigenstates of a particle bound in $\lambda z^{2m}$  $(\lambda >0, m=1,2,...)$ potentials is developed. It is represented by  ensembles of classical trajectories with energy  distributions that can take on negative values, for $m>1$ have integrable singularities at zero energy  and whose  mean energies coincide with quantum eigenenergies. The corresponding Schr\"odinger equation in classical representation is  analyzed.
\end{abstract}

\section{Introduction}\label{sec1}

Various aspects of relationship between classical and quantum mechanics are commonly explored by studying subjects such as: semiclassical quantization, limiting process $\hbar \to 0$, WKB approximation, measurement problem, quantum chaos, decoherence and alike (see, for example review \cite{Lan07}). An intriguing interplay between classical and quantum mechanics was introduced in the formulation of the {\it classical representation} of quantum states in 1D \cite{Sol93}. Stationary quantum states are represented by  ensembles of classical trajectories with energy distributions whose mean values are equal to quantum  eigenenergies. However, being of non-classical origin these distributions can take negative values and therefore, similarly to Wigner functions, belong to class of quasiprobability distributions. Nevertheless  we will refer to them in this work simply as to energy distributions. They are related to quantum position distributions via Abel transform and are  solutions of the Schr\"odinger equation in classical representation. The latter  takes  form of an integrodifferential equation and can be interpreted as  a balance equation with respect to some virtual exchange between classical states with energies $\varepsilon$ and $\tilde\varepsilon$ due to sub-barrier penetration \cite{Sol93}.  This interpretation has a profound connection with the treatment of quantum physics as the theory of {\it the global information field} \cite{Sol17}. Some solvable problems have been considered in  classical representation: harmonic oscillator, V-shaped potential $\lambda\mid z\mid$, time-dependent Feynman model for inelastic transitions in harmonic oscillator driven by  time dependent external force \cite{Sol93,Sol19} and  restricted model of s-states of  hydrogen atom \cite{Gro19}. In addition,  semiclassiacal approximation in  classical representation was formulated \cite{Sol20}.

In the present work we study the classical representation for oscillators with potentials $\lambda z^{2m}$. These potentials belong to wider  family of polynomial potentials which have been widely used in various aspects of solving the corresponding 1D Schr\"odinger equation: perturbation expansions \cite{Wen96}, high-order WKB solutions \cite{Kri77,Ben77,Rob00}, high-precision numerical solutions \cite{Mus11,Gau15} and exact solutions \cite{Bran13,Don19}. Additional motivation for studying these potentials is due to the fact that in the limit $m\to +\infty$ they tend to a square well with infinite (impenetrable) walls. As mentioned above the concept of classical representation relies on peculiar treatment of quantum tunneling \cite{Sol93}. Tunneling is absent in the case of infinite square well and a direct construction of classical representation in that case is problematic. We shall explore if some clarification can be gathered by studying the evolution of the classical representations for $\lambda z^{2m}$ potentials in the $m\to +\infty$ process. We mention that similar problems occur in construction of classical representations for other systems without tunneling such as 2D rigid rotor of treatment of azimuthal ($\varphi$-dependent) degree of freedom in 3D central potential problem.

The plan of the article is as follows. In Sect. \ref{sec2} we present  results of  numerical calculations of  quantum position distributions and eigenenergies of eigenstates. Section \ref{sec3} introduces  Abel transform which relates the quantum position distributions with classical energy  distributions. In Sect. \ref{sec4}  we present and discuss numerically obtained energy distributions of classical ensembles corresponding to ground and  excited states in various $x^{2m}$ potentials . Section \ref{sec5} contains the discussion of the form and peculiarities of the Schr\"odinger equation in classical representation for $x^{2m}$ potentials. Our conclusions are given in Sect. \ref{sec6}.

\section{Quantum position probability distributions}\label{sec2}

Schr\"odinger equation for stationary states $ \phi_n(z)$  $( -\infty<z<+\infty, n=0,1,2,...)$ of a particle of mass $\mu$ in potential  $\lambda z^{2m}$  $(\lambda >0, m=1,2,...)$ is
\begin{equation}
\frac{d^2\phi_n(z)}{dz^2}+\frac{2\mu}{\hbar^2}
[E_n-\lambda z^{2m}]\phi_n(z)=0.
\label{SE1}
\end{equation}
Introducing new independent variable $x$ and new spectral parameter $\varepsilon_n$:
\begin{equation}
z=\beta x,\;\;\;E_n=\gamma\varepsilon_n,\;\;\;\beta=\left(\frac{\hbar^2}{2\mu\lambda}\right)^{\frac{1}{2m+2}},\;\;\;\gamma=\lambda\beta^{2m},
\label{scal}
\end{equation}
(\ref{SE1}) is transformed into
\begin{equation}
\frac{d^2\psi_n(x)}{dx^2}+
[ \varepsilon_n -x^{2m}]\psi_n(x)=0.
\label{SE2}
\end{equation}
Once we solve for normalized eigenfunctions $\psi_n(x)$ and position probability distributions $\rho_n(x)=\psi^2_n(x)$  , the normalized eigenfunctions $\phi_n(z)$ and position probability distributions  $\tilde\rho_n(z)=\phi^2_n(z)$ are related to them by
\begin{equation}
\phi_n(z)=\frac{\psi_n(z/\beta)}{\sqrt{\beta}}, \;\;\;\tilde\rho_n(z)=\frac{\rho_n(z/\beta)}{\beta}.
\label{psirho}
\end{equation}

Therefore, in the rest of the paper we shall address only the eigenvalue problem (\ref{SE2}). The case $m=1$ (harmonic oscillator) has well known solutions
\begin{equation}
\rho_n(x)=\frac{1}{\sqrt{\pi}2^nn!}e^{-x^2} H^2_n(x), \;\;\;\varepsilon_n=2n+1,
\label{harosc}
\end{equation}
where $H_n(x)$ are Hermite polynomials. The case $m\to +\infty$ is also analytically solvable, because then  the potential $x^{2m}$ corresponds to square well with infinitely high walls located at $x=-1$ and $x=+1$. The solutions are
\begin{equation}
\rho_n(x)=\sin^2[\pi(n+1) (x+1)/2],\;\;\;\varepsilon_n=\frac{\pi^2}{4}(n+1)^2.
\label{infwell}
\end{equation}
In all other cases the eigenvalue problem (\ref{SE2}) has to be solved numerically.

\begin{figure}[htb]
\centering\resizebox{0.75\columnwidth}{!}{
\includegraphics{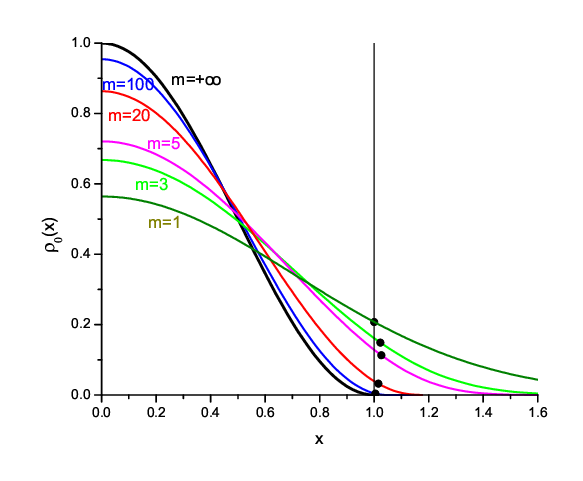}
}
\caption{ Ground-state position probability distributions $\rho_0(x)=\rho_0(-x)$  in various confining potentials $x^{2m}$. The values of distributions at classical turning points are marked by black points. Vertical thin line indicates the position of the infinite wall of the limiting ($m\to+\infty$) square well potential.}
\label{Fig1}
\end{figure}

In our calculations we have used a very efficient double exponential Sinc collocation method \cite{Gau15}. Figure \ref{Fig1} shows position probability distributions for the ground states ($n=0$) of various confining potentials $x^{2m}$. The black points define values of distributions at classical turning points $x_{tp}=\varepsilon_0^{1/(2m)}$. As $m$ increases the distributions are "pushed" towards the region $\mid x \mid<1$ ending up with $\rho_0(x)$ from (\ref{infwell}).

\begin{figure}[htb]
\centering\resizebox{0.75\columnwidth}{!}{
\includegraphics{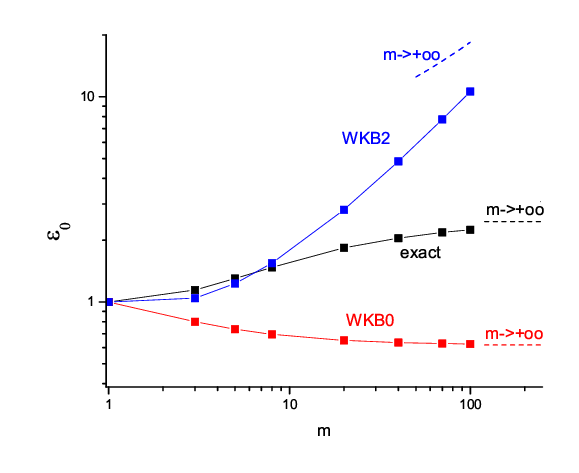}}
\caption{Ground state eigenvalues $\varepsilon_0$ for various confining potentials $x^{2m}$. Numerical results are represented by  black symbols (curve labeled exact). Zeroth and second order WKB approximations (curves labeled WKB0 and WKB2) are represented by red and blue symbols. Asymptotes corresponding to $m\to \infty$ are represented by dashed lines. } \label{Fig2}
\end{figure}

In Fig. \ref{Fig2} the curve labeled "exact" shows ground-state eigenvalues $\varepsilon_0$ for various values of $m$. The asymptotic approach to $m=+\infty$ value $\varepsilon_0$=$\pi^2/8$ from (\ref{infwell}) (the thin horizontal black dashed line) is indicated.

For the sake of illustrative comparison we  have also calculated eigenvalues using the WKB approximation. Explicit expressions for the ordinary (zeroth order) $\varepsilon_n^0$ and the second order $\varepsilon_n^2$ (which includes the correction of order  $\hbar^2$) WKB  approximations  have been derived in \cite{Kri77}:
\begin{equation}
\left(\varepsilon_n^{0}\right)^{\frac{m+1}{2m}}=\frac{\pi m}{B_1}(n+1/2),
\label{WKB0}
\end{equation}
\begin{equation}
\left(\varepsilon_n^2\right)^{\frac{m+1}{2m}}=\frac{\pi m}{2B_1}\left[n+1/2+\left((n+1/2)^2+\frac{B_1B_2(2m-1)(m-1)}{6\pi^2m^2}\right)^{1/2}\right],
\label{WKB2}
\end{equation}
\begin{equation}
B_1=B\left(\frac{1}{2m},\frac{3}{2}\right),\;\;\;B_2=B\left(1-\frac{1}{2m},\frac{1}{2}\right)
\label{B1B2}
\end{equation}
and $B(p,q)=\Gamma(p)\Gamma(q)/\Gamma(p+q)$ is the beta function.  In the case of harmonic oscillator (m=1) we find $ \varepsilon_n^0=\varepsilon_n^2=\varepsilon_n=2n+1$. If we let $m\to+\infty$ in (\ref{WKB0}) and (\ref{WKB2}) we find
\begin{equation}
\varepsilon_n^0\to\frac{\pi^2}{4}(n+1/2)^2,
\label{WKB0a}
\end{equation}
\begin{equation}
\varepsilon_n^{2}\to\frac{\pi^2}{16}\left\{n+1/2+\left[(n+1/2)^2+\frac{4m}{3\pi^2}\right]^{1/2}\right\}^2,
\label{WKB2a}
\end{equation}
so that $\varepsilon_n^0$ tends to constant value (\ref{WKB0a}) (which for large $n$ is consistent with (\ref{infwell})) while $\varepsilon_n^2$ diverges for large $m$.

For the ground state, as seen from Fig.\ref{Fig2} the $\varepsilon_0^0$ (curve labeled WKB0) as expected performs bad.  On the other hand, the $\varepsilon_0^{2}$ (curve labeled WKB2) provides reasonable predictions until $m\approx 8$ when it deviates toward the asymptote (\ref{WKB2a}) which is shown as dashed blue curve.

\begin{figure}[htb]
\centering\resizebox{0.75\columnwidth}{!}{
\includegraphics{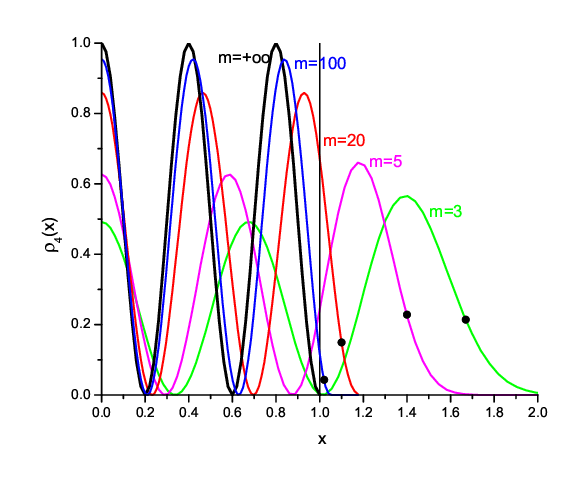}}
\caption{ Same as Fig.\ref{Fig1} but for excited $n=4$ state.} \label{Fig3}
\end{figure}

As a representative of excited states we shall take the $n=4$ state. Figure \ref{Fig3} shows quantum position densities for $n=4$ states for various $x^{2m}$ potentials. With the increase of $m$ distributions are pushed towards the region $-1<x<1$ approaching asymptotically the distribution (\ref{infwell}) of the paticle in the box with infinite walls.

\begin{figure}[htb]
\centering\resizebox{0.75\columnwidth}{!}{
\includegraphics{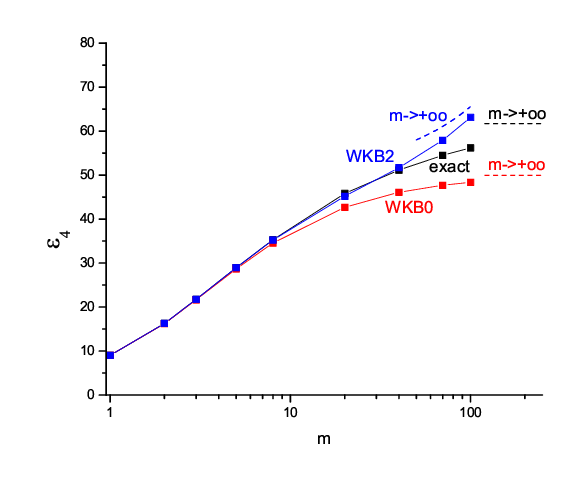}}
\caption{Same as Fig\ref{Fig2} but for $n=4$ states. } \label{Fig4}
\end{figure}

The eigenvalues $\varepsilon_4$ as a function of $m$ are shown in Figure \ref{Fig4} as black points and the corresponding asymptotic value (\ref{infwell}) as dashed black line. The zeroth order WKB approximation $\varepsilon_4^0$  (curve labeled WKB0) performs much better than in the case of ground state, but somewhere arround $m=8$ departs from the vicinity of the  exact results and eventually approach its asymtote (\ref{WKB0a}) (red dashed line). The second order approximation $\varepsilon_4^2$  (curve labeled WKB2) performes even better and for larger $m$ but also departs from the vicinity of the exact values around $m=40$ and approach its asymptote (\ref{WKB2a}) (dashed blue curve) .

The zeroth and second order WKB approximations have been used here for illustrative purposes. It should be mentioned that  elaborate techniques exist for calculating  very high order WKB approximations related to $x^{2m}$ potentials, which are very accurate, expecially for exited states \cite{Ben77,Rob00}.

\section{ Classical representation and Abel transform}\label{sec3}

For the class of symmetric potentials $V(-x)=V(x)$, $V(0)=0$ that monotonically increase for $x>0$ the detailed procedure for construction of classical representation was developed in \cite{Sol93}. Since the potential $v(x)=x^{2m}$ belongs to this class we closely follow that procedure.

An eigenstate  with position distribution $\rho_n(x)$
is represented as an ensemble of classical trajectories
oscillating between the turning points $-x_{tp}(\varepsilon)$ and  $x_{tp}(\varepsilon)=\varepsilon^{1/(2m)}$ and
distributed over the energies $\varepsilon$ in classically allowed region $\varepsilon>v(x)=x^{2m}$ with a certain  distribution $f_n(\varepsilon)$:
\begin{equation}
\rho_n(x)=\int_{x^{2m}}^{+\infty}q(\varepsilon,x)f_n(\varepsilon) d\varepsilon,
\label{rhof}
\end{equation}
where the $x$-probability distribution for the classical ensemble $q(\varepsilon,x)$ at energy $\varepsilon$ is proportional to the time spent by particle in the interval $dx$:
\begin{equation}
q(\varepsilon,x)=\frac{2}{T(\varepsilon)}\frac{dt}{dx}=
\frac{1}{T(\varepsilon)\sqrt{\varepsilon-x^{2m}}},\label{q}
\end{equation}
and $T(\varepsilon)$ is classical period
\begin{equation}
T(\varepsilon)=\int_{-\varepsilon^{\frac{1}{2m}}}^{\varepsilon^{\frac{1}{2m}}}\frac{dx}{\sqrt{\varepsilon-x^{2m}}}=\frac{1}{m}B\left(\frac{1}{2m},\frac{1}{2}\right)\varepsilon^\frac{1-m}{2m}, \label{T}
\end{equation}
so that  normalization condition holds
\begin{equation}
\int_{-\varepsilon^{\frac{1}{2m}}}^{\varepsilon^{\frac{1}{2m}}}q(\varepsilon,x)dx=1. \label{nq}
\end{equation}
The classical relations (\ref{q}) and (\ref{T}) hold because the classical hamiltonian related to Schr\"odinger eqation (\ref{SE1}) is
\begin{equation}
H_z^{cl}=\frac{p_z^2}{2\mu}+\lambda z^{2m},
\label{Hz}
\end{equation}
and after applying the change of variables (\ref{scal}) the classical hamiltonian related to Schr\"odinger eqation (\ref{SE2}) is
\begin{equation}
H_x^{cl}=p_x^2+x^{2m},\;\;\; p_x=p_z/\sqrt{2\mu \gamma},
\label{Hx}
\end{equation}
so that the equation of motion is
\begin{equation}
\frac{dx}{dt}=\frac{\partial H_x^{cl}}{\partial p_x}=2p_x=2\sqrt{\varepsilon-x^{2m}}.
\label{eqm}
\end{equation}

Due to the condition
(\ref{nq}) the integration of (\ref{rhof}) over $x$ leads to the
normalization condition of the  distribution $f_n(\varepsilon)$:
\begin{equation}
\int_0^{+\infty}f_n(\varepsilon)d\varepsilon= \int_{-\infty}^{+\infty}\rho_n(x)
dx=1.\label{nf}
\end{equation}

We can rewrite (\ref{rhof}) as
\begin{equation}
\rho_n(x)=\int_v^{+\infty}\frac{f_n(\varepsilon)/T(\varepsilon)}{\sqrt{\varepsilon-v}}d\varepsilon.
\label{Abel1}
\end{equation}
It is the well known Abel transform with respect to
variable $v$. Its reverse expression reads \cite{Gor91}
\begin{equation}
f_n(\varepsilon)
=-\frac{T(\varepsilon)}{\pi} \int_{\varepsilon}^{+\infty}
\frac{d\rho_n(x(v))}{dv}\frac{dv}{\sqrt{v-\varepsilon}},
\label{Abel2}
\end{equation}
where $x(v)=v^{\frac{1}{2m}}$ is the inverse function to  $v(x)=x^{2m}$.

The transforms (\ref{Abel1}), (\ref{Abel2}) can be understood, in some broader sense, as connecting two representations of the same state \cite{Sol93}. The representation defined by $f_n(\varepsilon)$  was called "classical representation" since the kernel $q(\varepsilon,x)$ is the classical probability density and $f_n(\varepsilon)$ has the meaning of the
energy distribution in the classical ensemble. These distributions, as we shall see, can take negative values and therefore are quasiprobabilty distributions.

In addition, an important relation holds \cite{Sol93}
\begin{equation}
\int_0^{+\infty}\varepsilon f_n(\varepsilon)d\varepsilon=\varepsilon_n,\label{ave}
\end{equation}
meaning that average energy of classical
ensemble is equal to eigenenergy of quantum state .

\section{Energy distributions $f_n(\varepsilon)$}\label{sec4}

In the case of harmonic oscillator ($m=1)$  $f_n(\varepsilon)$ distributions are known analytically \cite{Sol93}(see also section \ref{sec5} below)
\begin{equation}
f_n(\varepsilon)=(-1)^ne^{-\varepsilon}L_n(2\varepsilon),
\label{fnm1}
\end{equation}
where $ L_n(x)$ is Laguerre polynomial.

\begin{figure}[htb]
\centering\resizebox{0.75\columnwidth}{!}{
\includegraphics{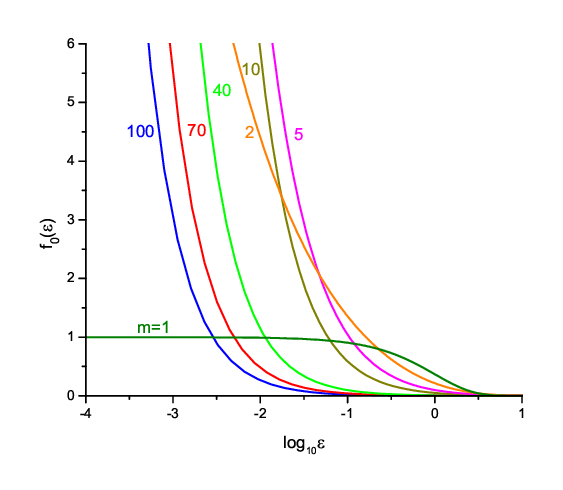}}
\caption{Ground-state  energy distributions $f_0(\varepsilon)$ for various $x^{2m}$ potentials.} \label{Fig5}
\end{figure}

\begin{figure}[htb]
\centering\resizebox{0.75\columnwidth}{!}{
\includegraphics{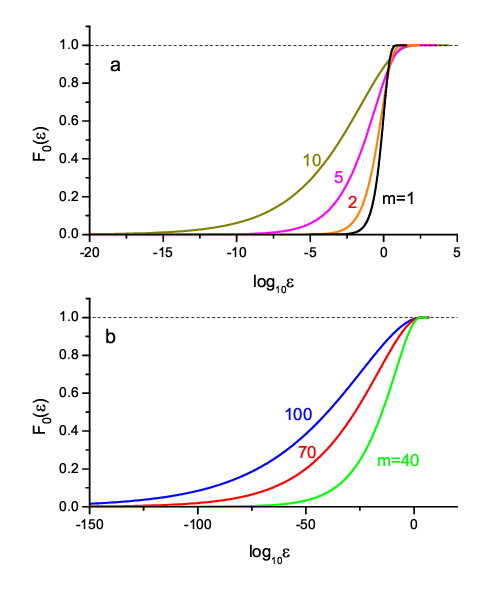}}
\caption{Ground-state  cumulative energy distributions $F_0(\varepsilon)$ for various $x^{2m}$ potentials.} \label{Fig6}
\end{figure}

We used  numerically  calculated quantum distributions $\rho_n(x)$ and numerical differentiation and integration in (\ref{Abel2}) to determine  $f_n(\varepsilon)$  for various quantum states and  $x^{2m}$ potentials. Figure \ref{Fig5} shows ground-state distributions $f_0(\varepsilon)$ for a number of potentials with $m=1$ to $m=100$. We see that in the limit $\varepsilon \to 0$ only in the case of harmonic oscillator ($m=1$) $f_0(\varepsilon)$ remains finite (in accord with (\ref{fnm1}),  $f_0(0)=1$)  while in all other cases they diverge to $+\infty$. Closer inspection of the integral (\ref{Abel2}) for $\varepsilon \to 0$ shows that the behavior  of $f_n(\varepsilon)$ distributions is given by (see Appendix)
\begin{equation}
\begin{array}{lll}f_n(0) =(-1)^n & m=1, \\
f_n(\varepsilon)\approx\frac{(-1)^n}{2\pi}\mid c_{n1}\mid B(\frac{1}{4},\frac{1}{2})\log(\varepsilon^{-\frac{1}{4}})\varepsilon^{-\frac{1}{4}}&m=2,\\
f_n(\varepsilon)\approx\frac{(-1)^n}{m\pi}\mid c_{n1}\mid B(\frac{1}{2m},\frac{1}{2})S_{m1}\varepsilon^{-1+\frac{3}{2m}}&m>2,\\
c_{n1}= \frac{d^2\rho_n(x)}{dx^2} \mid_{x=0},\\
S_{m1}=\sum_{k=0}^\infty\frac{(2k-1)!!}{(2k)!![(2k+1)m-2]}.
\end{array}
\label{sing}
\end{equation}

Note that indeed in Fig. \ref{Fig5} the slope of rising of $f_0(\varepsilon)$ for the quartic potential $(m=2)$ is different from other cases. The singularities (\ref{sing}) are integrable so that {\it cumulative} energy distributions  are well defined:
\begin{equation}
F_n(\varepsilon)=\int_0^{\varepsilon} f_n(\tilde\varepsilon)d\tilde\varepsilon.
\label{Feps}
\end{equation}
However, in the limit $m\to +\infty$ the singularity is of the type $\varepsilon^{-1}$, that is nonintegrable, so that the existence of classical representation for the infinite square well is problematic.

Cumulative distributions corresponding to cases shown in Fig. \ref{Fig5} are shown in Fig. \ref{Fig6}. We can see that as $m$ increases the values of $\varepsilon$ where  $F_0(\varepsilon)$ distributions are non negligible are getting closer to  $\varepsilon=0$. Results shown in Fig. \ref{Fig6} confirm the normalization condition (\ref{nf}). In addition, we have verified by numerical integration that condition (\ref{ave}) is fulfilled in all cases.

Note that $f_0(\varepsilon)>0$ and  $F_0(\varepsilon)>0$ for all values of $m$ and therefore they are not quasiprobability distributions but rather ordinary probability distributions. Apparently, this occurs because the ground-state quantum position distributions $\rho_0(x)$ are nodeless. This will not be the case with excited states.

\begin{figure}[htb]
\centering\resizebox{0.75\columnwidth}{!}{
\includegraphics{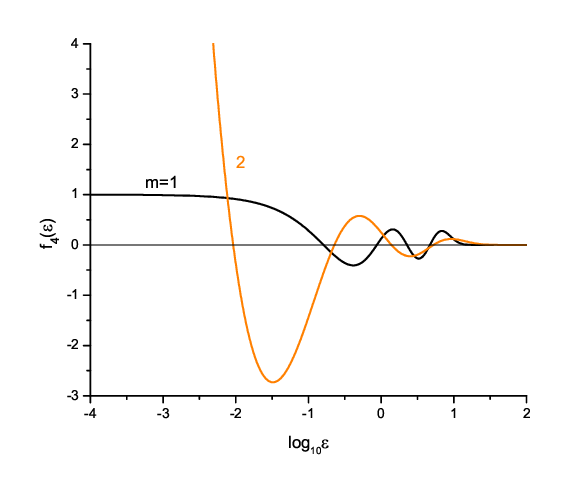}}
\caption{Excited-state ($n=4$) energy distributions $f_4(\varepsilon)$ for harmonic ($m=1$) and quartic ($m=2$) oscillators.} \label{Fig7}
\end{figure}

\begin{figure}[htb]
\centering\resizebox{0.75\columnwidth}{!}{
\includegraphics{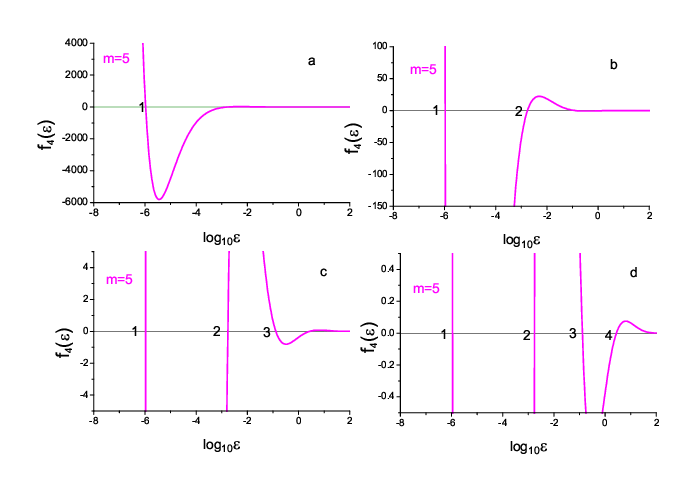}}
\caption{Same as Fig. \ref{Fig7} but for $m=5$ case and on four different scales of $f_4(\varepsilon)$. Points labeled with 1-4 are nodes of the $f_4(\varepsilon)$ distribution. } \label{Fig8}
\end{figure}

\begin{figure}[htb]
\centering\resizebox{0.75\columnwidth}{!}{
\includegraphics{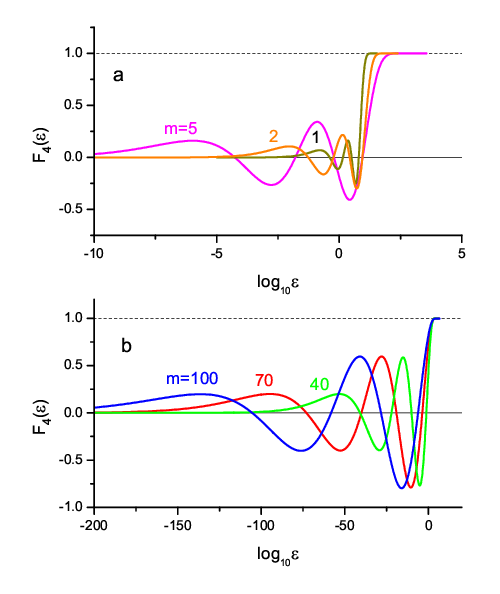}}
\caption{Excited-state ($n=4$)  cumulative energy distributions $F_4(\varepsilon)$ for various $x^{2m}$ potentials.} \label{Fig9}
\end{figure}

\begin{figure}[htb]
\centering\resizebox{0.75\columnwidth}{!}{
\includegraphics{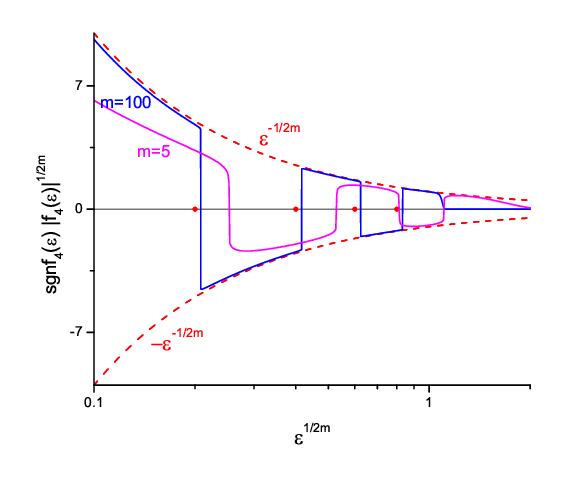}}
\caption{Scaled energy distributions  ${\rm sgn} f_4(\varepsilon)\mid f_4(\varepsilon)\mid^{\frac{1}{2m}}$ as functions of  $\varepsilon^{\frac{1}{2m}}$ for the cases $m=5$ and $m=100$. The red pints are nodes (\ref{nodes}) of the limiting distribution for $m\to +\infty$.} \label{Fig10}
\end{figure}

In Fig. \ref{Fig7} we  present  results for $n=4$ excited states of  harmonic ($m=1$) and quartic ($m=2$) oscillators, confirming the behavior of  $f_4(\varepsilon)$ distributions as predicted by (\ref{sing}).  As $m$ increases the amplitudes of oscillations of $f_4(\varepsilon)$ enormously increase so that it is not possible the represent all details on a single figure. Figure \ref{Fig8} shows four different scales which are necessary to resolve all four nodes of $f_4(\varepsilon)$ distribution in the $m=5$ case. Cumulative energy distributions are better adopted for compact presentation. Figure \ref{Fig9}a shows  $F_4(\varepsilon)$ distributions corresponding to lower values of $m$ and Fig. \ref{Fig9}b those  corresponding to higher values of $m$. In the cases of extremely high values of $m$ the regions where the $F_4(\varepsilon)$ distributions are not negligible are extremly close to $\varepsilon=0$. With the increase of $m$ the nodes of distributions move toward  $\varepsilon=0$. These results also confirm the normalization condition (\ref{nf}) and we have verified by numerical integration that relation (\ref{ave}) holds in all cases.

We have chosen as examples two symmetric states ($n=0$ and $n=4$) and studied them for various values of $m$. In the case of antisymmetric states (odd $n$) all properties of  $f_n(\varepsilon)$ and  $F_n(\varepsilon)$  distributions remain the same with the only difference that, according to (\ref{sing}), for $m>1$ the $f_n(\varepsilon)$ distributions diverge to $-\infty$ when $\varepsilon\to 0$.

The  energy distributions $\tilde f_n(E)$ and $\tilde F_n(E)$ corresponding  to $V(z)=\lambda z^{2m}$ potential are related to $f_n(\varepsilon)$ and  $F_n(\varepsilon)$ distributions through the simple relations
\begin{equation}
 \tilde f_n(E)=f_n(E/\gamma)/\gamma, \;\;\;\tilde F_n(E)=F_n(E/\gamma).
\label{fnEFnE}
\end{equation}

Considering  the behavior of energy distributions $f_n(\varepsilon)$ in the limit of $m\to +\infty$, some additional information can be obtained with an appropriate scaling of both $ f_n(\varepsilon)$  and $\varepsilon$. This is shown in  Fig. \ref{Fig10} where the scaled distributions ${\rm sgn} f_4(\varepsilon)\mid f_4(\varepsilon)\mid^{\frac{1}{2m}}$ (with ${\rm sgn} f\equiv f/\mid f\mid$) as functions of  $\varepsilon^{\frac{1}{2m}}$ are shown for the cases $m=5$ and $m=100$. Note that in the $m=5$ case the complete behavior of the scaled distribution is shown on the single graph, while for the unscaled quantities we needed four graphs, as seen from Fig. \ref{Fig8}.

The result for the $m=100$ case  in  Fig. \ref{Fig10} suggests the following limiting form for the scaled energy distribution  ${\rm sgn} f_n(\varepsilon)\mid f_n(\varepsilon)\mid^{\frac{1}{2m}}$  when $m\to +\infty$: For $\varepsilon^{\frac{1}{2m}}<1$, it consists of alternating arcs which lie on $\varepsilon^{-\frac{1}{2m}}$ and $-\varepsilon^{-\frac{1}{2m}}$ curves with jumps at the nodes of the distribution defined by the expression
\begin{equation}
\varepsilon^{\frac{1}{2m}}_k=\frac{k}{n+1},\;\;k=1,2,...n.
\label{nodes}
\end{equation}
For $\varepsilon^{\frac{1}{2m}}\approx 1$ there is abrupt fall of the distribution to zero. We have verified that this pattern is correct for other values of $n$. However, the emerging forms of distributions $f_n(\varepsilon)$ all diverge as $\pm 1/\varepsilon$ as $\varepsilon \to 0$, in accord with (\ref{sing}) and are therefore nonintegrable. Thus, the energy distributions for the infinite square well are not well defined, although for any potential $x^{2m}$ with large but finite $m$ they are well defined.

\begin{figure}[ht]
 \centerline{\includegraphics[width=16cm]{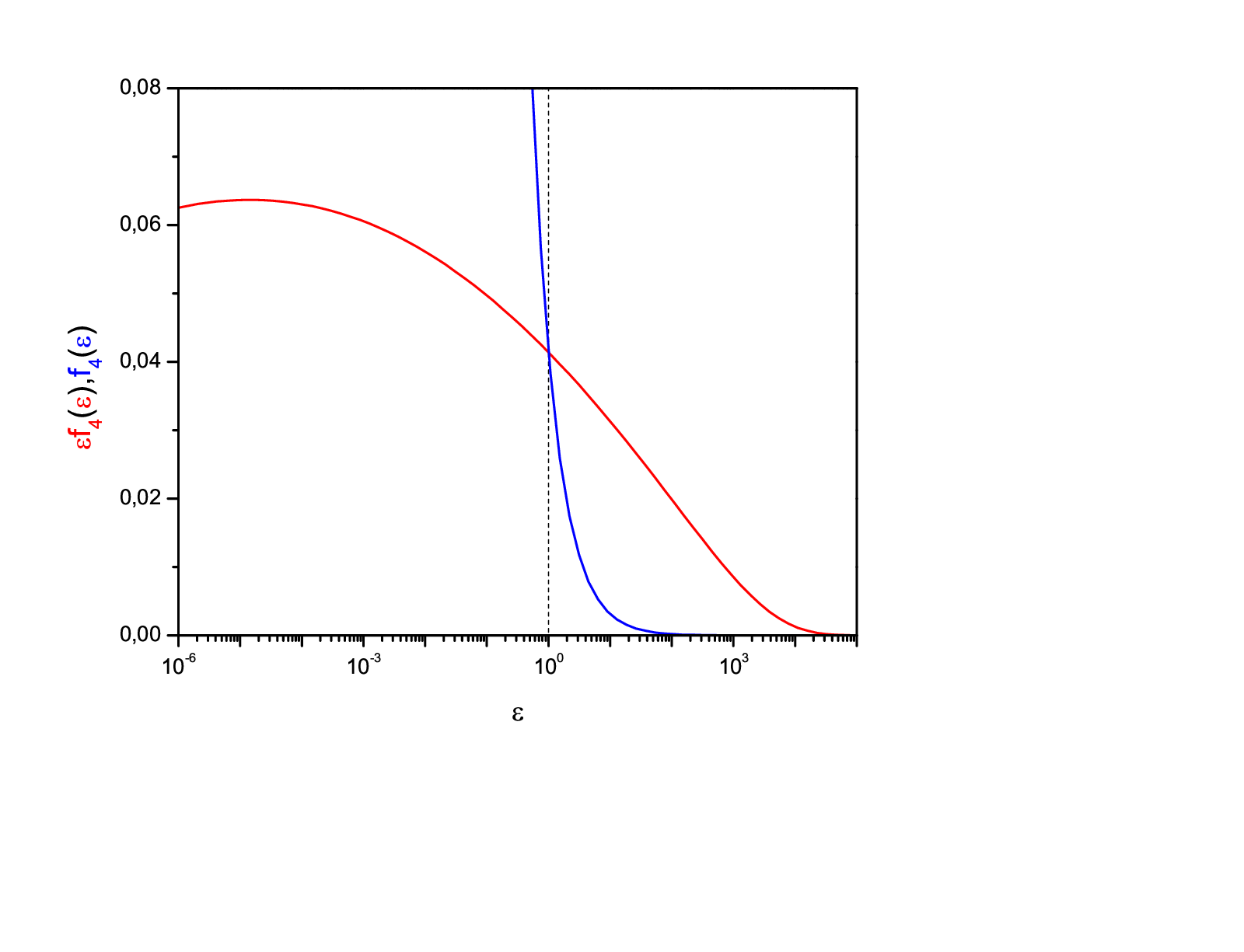}}
 \caption{Distribution   $f_4(\varepsilon)$ (blue curve) and $\varepsilon f_4(\varepsilon)$ (red curve) for  $m=100$ as functions of  $\varepsilon$ in logarithmic scale.} \label{Fig11}
 \end{figure}

Thus the eigenstates of rectangular well in classical representation are new class of generalized functions which are located at the point $\varepsilon=0$ and have complicate structure providing orthonormalization conditions. For ground state, which satisfies normalization condition
\begin{equation}
\int_0^{ 1} f_0^{\rm app.}(\varepsilon)d\varepsilon=1,\label{norm}
\end{equation}
it is one of the possible representation of delta function $\delta(x)$.

Figure \ref{Fig10} provokes the wrong impression that the distribution $f_n(\varepsilon)$ is negligible for $\varepsilon>1$. In fact, it depends on what it is used for. For large $m$, there is a tail of the distribution in the region $\varepsilon>1$, which, for example, gives a dominant contribution to the calculation of the mean value of $\varepsilon$, which should coincide with the eigenvalue $E_n$. In Fig.\ref{Fig11} $f_4(\varepsilon)$ and $\varepsilon f_4(\varepsilon)$ are shown for $n=4$ and $m=100$ in a logarithmic scale, and the mean value
\begin{equation}
\bar{\varepsilon}=\int_0^{ \infty} \varepsilon f_4(\varepsilon)d\varepsilon
\nonumber
\end{equation}
coincides with the eigenenergy $E_4= 56.17$. As can be seen from Fig.\ref{Fig11}, the region $\varepsilon\le 1$ contributes to $\bar{\varepsilon}$ less than 0.06 (or 0.1\% ) .

\section{ Schr\"odinger equation in classical representation}\label{sec5}

It was shown in \cite{Sol93} that starting from the Schr\"odinger equation, in our case (\ref{SE2}), one can derive the linear third-order differential equation for the position probability density $\rho(x)=\psi(x)^2$
\begin{equation}
\frac{d^3\rho_n(x)}{dx^3}-
4[v(x)-\varepsilon_n]\frac{d\rho_n(x)}{dx}-2\frac{dv(x)}{dx}\rho_n(x)=0,
\label{SE3}
\end{equation}
where $v(x)=x^{2m}$. Introducing function
\begin{equation}
\varphi_n(\varepsilon)=\frac{f_n(\varepsilon)}{T(\varepsilon)}
\label{varphi}
\end{equation}
and substituting  (\ref{Abel1}) in (\ref{SE3}) one arrives at integrodifferential equation \cite{Sol93}
\begin{equation}
(\varepsilon - \varepsilon_n)\varphi_n(\varepsilon)=\frac{2}{15\pi}\int_\varepsilon^{+\infty} Q(\tilde\varepsilon,\varepsilon)\frac{d^3\varphi_n(\tilde\varepsilon)}{d\tilde\varepsilon^3}d\tilde\varepsilon
\label{Inteq}
\end{equation}
with the kernel
\begin{equation}
Q(\tilde\varepsilon,\varepsilon)=\int_{\varepsilon^{\frac{1}{2m}}}^{\tilde\varepsilon^{\frac{1}{2m}}}(x^{2m}-\varepsilon)^{-1/2}\frac{d^3}{dx^3}(\tilde\varepsilon-x^{2m})^{5/2}dx.
\label{kernel1}
\end{equation}
The boundary conditions associated with eigevalue problem (\ref{Inteq}) are  for $\varepsilon \to +\infty$
\begin{equation}
\varphi_n(\varepsilon)\to 0
\label{bcinf}
\end{equation}
and for $\varepsilon\to 0$ (see Appendix)
\begin{equation}
\begin{array}{lll}\varphi_n(0) =\frac{(-1)^n}{\pi},& m=1, \\
\varphi_n(\varepsilon)\approx\frac{(-1)^n}{\pi}\mid c_{n1}\mid\log(\varepsilon^{-\frac{1}{4}}),&m=2,\\
\varphi_n(\varepsilon)\approx\frac{(-1)^n}{\pi}\mid c_{n1}\mid S_{m1}\varepsilon^{-\frac{1}{2}+\frac{1}{m}},&m>2,\\
c_{n1}= \frac{d^2\rho_n(x)}{dx^2} \mid_{x=0},\\
S_{m1}=\sum_{k=0}^\infty\frac{(2k-1)!!}{(2k)!![(2k+1)m-2]},
\end{array}
\label{bc0}
\end{equation}
which is in accord with (\ref{T}), (\ref{sing}) and (\ref{varphi}).

Performing the integration in (\ref{kernel1}) one finds
\begin{equation}
Q(\tilde\varepsilon,\varepsilon)=\sum_{k=1}^3c_k(m)I_k,
\label{kernel2}
\end{equation}
\begin{equation}
c_1(m)=-\frac{15m^2}{2},\;\;c_2(m)=\frac{45m(2m-1)}{2},\;\;c_3(m)=-5(2m-1)(m-1),
\label{ckm}
\end{equation}
\begin{equation}
I_k=\int_{\varepsilon}^{\tilde\varepsilon}(v-\varepsilon)^{-\frac{1}{2}}(\tilde\varepsilon-v)^{-\frac{3}{2}+k}v^{3-k-\frac{1}{m}}dv
\nonumber
\end{equation}
\begin{equation}
=\varepsilon^{3-k-\frac{1}{m}} (\tilde\varepsilon-\varepsilon)^{k-1}B\left(k-\frac{1}{2},\frac{1}{2}\right) \;F\left(k-3+\frac{1}{m},\frac{1}{2};k;-\frac{\tilde\varepsilon-\varepsilon}{\varepsilon}\right)
\label{I123}
\end{equation}
where $F(a,b;c;z)$ is hypergeometric function and the integral was taken from \cite{GR}.

Obviously, solving numerically integral equation (\ref{Inteq}) with the kernel (\ref{kernel2}) would be a challenging task. The direct solution of Abel transform (\ref{Abel2}) which we demonstrated in the previous section seems to be more efficient method.

One exception is the case (m=1) of harmonic oscillator. In that case only the first two terms in (\ref{kernel2}) survive, the corresponding two  hypergeometric functions in (\ref{I123}) reduce to polynomials and we find
\begin{equation}
Q(\tilde\varepsilon,\varepsilon)=\frac{15\pi}{2}(\tilde\varepsilon-2\varepsilon).
\label{kernelho}
\end{equation}
Substituting (\ref{kernelho}) in (\ref{Inteq}) after partial integration the following differential equation is obtained
\begin{equation}
(\varepsilon - \varepsilon_n)\varphi_n(\varepsilon)=\varepsilon\frac{d^2\varphi_n(\varepsilon)}{d\varepsilon^2}+\frac{d\varphi_n(\varepsilon)}{d\varepsilon}.
\label{difeqho}
\end{equation}
Its solution obeying the boundary conditions (\ref{bcinf}) and (\ref{bc0}) is
\begin{equation}
\varphi_n(\varepsilon)=\frac{(-1)^n}{\pi}e^{-\varepsilon}L_n(2\varepsilon), \;\;\; \varepsilon_n=2n+1,
\label{varphim1}
\end{equation}
which is in accord with  (\ref{varphi})  and  (\ref{fnm1}) since  $T(\varepsilon)=\pi$ for $m=1$.

In terms of the non-scaled variables, that is for potential $V(z)=\frac{1}{2}\mu \omega^2z^2$, we find from (\ref{scal}) $\gamma=\hbar\omega/2$ and therefore
\begin{equation}
\tilde f_n(E)=\frac{f_n(\frac{E}{\gamma})}{\gamma}=(-1)^n\frac{2}{\hbar \omega}e^{-\frac{2E}{\hbar \omega}}L_n\left(\frac{4E}{\hbar \omega}\right), \;\;\; E_n=\gamma\varepsilon_n=\hbar\omega(n+\frac{1}{2}),
\label{ffnm1}
\end{equation}
which is the result found in \cite{Sol93} except for missing the factor $(-1)^n$.

Interestingly, in the limit $m\to + \infty$, the hypergeometric functions in (\ref{I123}) also reduce to polynomials and we find
\begin{equation}
Q(\tilde\varepsilon,\varepsilon)=-\frac{15\pi m^2}{16}(\tilde\varepsilon^2-18\tilde\varepsilon \varepsilon+25\varepsilon^2),
\label{Qminf}
\end{equation}
which tends to infinity, confirming our earlier statement that the existance of classical representation for infinite square well is problematic.

\section{ Concluding remarks}\label{sec6}

We have presented a systematic study of classical representations for variety of potentials $\lambda z^{2m}$. The  energy  distributions $f_n(\varepsilon)$ were determined by direct numerical solution of the Abel integral equation (\ref{Abel2}).
It was found that, except for the $m=1$ case of harmonic oscillator, in all other cases the energy distributions diverge at $\varepsilon=0$. These singularities are however integrable and cumulative energy distributions are well defined. We have also numerically verified that the mean energies of $f_n(\varepsilon)$ distributions coincide with quantum eigenenergies $\varepsilon_n$.
Although quantum position densities and eigenvalues for $m\to +\infty$ tend towards those of the infinite square well  potential, this limit in the case of $f_n(\varepsilon)$ distributions lead to nonintegrable singularities indicating that classical representation in this case is problematic. This is also confirmed by analysis of the Shr\"odinger equation in classical representation for the $m\to +\infty$ case.

Further developments of the theory should address the general 1D potentials (beyond the symmetrical case) and extensions towards the multidimensional systems.

\section*{Acknowledgments}

This work was supported by Serbia-JINR collaboration program.

\section*{Appendix A}

\renewcommand{\theequation}{A.\arabic{equation}}
\setcounter{equation}{0}

Here we derive results presented in expressions (\ref{sing}).

The result for $m=1$ follows if we substitue $\varepsilon =0$ in (\ref{Abel2}) and use for $\rho_n(x)$  the expression (\ref{harosc}). Than we employ the relations
\begin{equation}
H_n'(x)=2nH_{n-1}(x),\;\;\;H_n(x)=2xH_{n-1}(x)-2(n-1)H_{n-2}(x)
\label{HH}
\end{equation}
to reduce (\ref{Abel2}) to $f_n(0)=-f_{n-1}(0)$ and since by direct calculation $f_0(0)=1$ the result (\ref{sing}),  $f_n(0)=(-1)^n$  follows.

In order to study the singular behavior of the Abel transform  (\ref{Abel2}) in the limit $\varepsilon\to 0$ and for $m>1$   we start from
\begin{equation}
\varphi_n(\varepsilon)=\frac{f_n(\varepsilon)}{T(\varepsilon)}
=-\frac{1}{\pi} \int_{\varepsilon}^{+\infty}
\frac{d\rho_n(x(v))}{dv}\frac{dv}{\sqrt{v-\varepsilon}}
\nonumber
\end{equation}
\begin{equation}
=-\frac{1}{\pi} \int_{\varepsilon}^{\delta}
\frac{d\rho_n(x(v))}{dv}\frac{dv}{\sqrt{v-\varepsilon}}-\frac{1}{\pi} \int_{\delta}^{+\infty}
\frac{d\rho_n(x(v))}{dv}\frac{dv}{\sqrt{v-\varepsilon}}
\nonumber
\end{equation}
\begin{equation}
=-\frac{1}{\pi} \int_{\varepsilon^{\frac{1}{2m}}}^{\delta^{\frac{1}{2m}}}
\frac{d\rho_n(x)}{dx}\frac{dx}{\sqrt{x^{2m}-\varepsilon}}+O(1)
\label{phin1}
\end{equation}
We chose $\delta$  such that $\varepsilon^{\frac{1}{2m}}<x<\delta^{\frac{1}{2m}}<<1$, and use power expansion
\begin{equation}
\frac{d\rho_n(x)}{dx}=\sum_{p=1}^{p_{\rm max}}c_{np}x^{2p-1}, \;\;
c_{np}=\frac{1}{(2p-1)!}\frac{d^{2p}\rho_n(x)}{dx^{2p}}\mid_{x=0},
\label{drhodx1}
\end{equation}
because from (\ref{SE2}) it follows  that
\begin{equation}
\frac{d^{2p-1}\rho_n(x)}{dx^{2p-1}}\mid_{x=0}=0.
\label{drhodx2pm1}
\end{equation}
The upper limit of summation $p_{\rm max}$ will be determined below.
We also substitute in (\ref{phin1}) the expansion
\begin{equation}
\frac{1}{\sqrt{x^{2m}-\varepsilon}}=\frac{1}{x^m\sqrt{1-\frac{\varepsilon}{{x^{2m}}}}}=\frac{1}{x^m}\sum_{k=0}^\infty\frac{(2k-1)!!\varepsilon^k}{(2k)!!x^{2mk}}
\label{sqrt}
\end{equation}
with convention $(-1)!!=0!!=1$. Therefore
\begin{equation}
\varphi_n(\varepsilon)\approx-\frac{1}{\pi}\sum_{p=1}^{p_{\rm max}}c_{np}\sum_{k=0}^\infty\frac{(2k-1)!!\varepsilon^k}{(2k)!!} \int_{\varepsilon^{\frac{1}{2m}}}^{\delta^{\frac{1}{2m}}}x^{2p-1-m-2mk}dx+O(1).
\label{phin2}
\end{equation}

For $m>2$ we find
\begin{equation}
\varphi_n(\varepsilon)\approx-\frac{1}{\pi}\sum_{p=1}^{p_{\rm max}}c_{np}S_{mp}\varepsilon^{-\frac{1}{2}+\frac{p}{m}}+O(1),
\label{phin3}
\end{equation}
where
\begin{equation}
S_{mp}=\sum_{k=0}^\infty\frac{(2k-1)!!}{(2k)!![(2k+1)m-2p]}
\label{sum}
\end{equation}
is according to {\it Raabe's convergence criterion} \cite{GR} a convergent series.

The derivation of  (\ref{phin3}) is valid under the assumption that $\varphi_n(\varepsilon)$ is a {\it  singular function} at $\varepsilon=0$, which puts the restriction on the exponent in (\ref{phin3}): $-\frac{1}{2}+\frac{p}{m}<0$. This sets the upper limits in summations:

\begin{equation}
p_{\rm max}=\left\{
\begin{array}{ll}\frac{m}{2}-1,& \;\; m-{\rm even} \\
\frac{m-1}{2},& \;\;m-{\rm odd}.
\end{array}
\right.
\label{pmax}
\end{equation}

For $m=2$ we have a special case because the integral in (\ref{phin2}) for $p=1$ and $k=0$ is the only one that  gives singular (logarithmic) term and we find
\begin{equation}
\varphi_n(\varepsilon)\approx-\frac{1}{\pi}c_{n1}\log(\varepsilon^{-\frac{1}{4}})+O(1).
\label{phin4}
\end{equation}

The corresponding behavior of energy distributions  follows from relation  $f_n(\varepsilon)=T(\varepsilon)\varphi_n(\varepsilon)$ with $T(\varepsilon)$ taken from (\ref{T}). For $m>2$ we find
\begin{equation}
f_n(\varepsilon)\approx-\frac{1}{m\pi}B\left(\frac{1}{2m},\frac{1}{2}\right)\sum_{p=1}^{p_{\rm max}}c_{np}S_{mp}\varepsilon^{-1+\frac{2p+1}{2m}},
\label{fnpasym}
\end{equation}
and for $m=2$
\begin{equation}
f_n(\varepsilon)\approx-\frac{1}{2\pi}c_{n1}B\left(\frac{1}{4},\frac{1}{2}\right)\varepsilon^{-\frac{1}{4}}\log(\varepsilon^{-\frac{1}{4}}).
\label{fn1asym}
\end{equation}

We note that by using (\ref{SE2}), all coefficients in the expansion (\ref{drhodx1}) can be calculated analytically:
\begin{equation}
c_{np}=\frac{(-4\varepsilon_n)^{p-1}}{(2p-1)!}c_{n1},
\label{cnpcn1}
\end{equation}
\begin{equation}
c_{n1}=\frac{d^2\rho_n(x)}{dx^2}\mid_{x=0}=\left\{
\begin{array}{ll}-2\varepsilon_n\rho_n(0),& \;\; n-{\rm even} \\
2\left(\frac{d\psi_n}{dx}\mid_{x=0}\right)^2, &\;\;n-{\rm odd}.
\end{array}
\right.
\label{cn1}
\end{equation}
The relation (\ref{cnpcn1}) is valid for $1\leq p \leq m$ and from (\ref{pmax}) we see that $p_{\rm max}<m$.

The expressions (\ref{sing}) and (\ref{bc0}) in the main text are obtained by keeping only the first terms ($p=1$) in the sums  (\ref{fnpasym}) and  (\ref{phin3}) and the property $c_{n1}=(-1)^{n+1}\mid c_{n1}\mid$ which follows from (\ref{cn1}).

\end{document}